\documentclass[10pt]{iopart}
\usepackage{iopams}
\usepackage{setstack}
\usepackage[english]{babel}
\usepackage[utf8]{inputenc}
\usepackage{cite}
\usepackage{graphicx}
\usepackage{dsfont} 

\newcommand{\Bop}[3]{\raisebox{-2pt}{\large{$\breve{A}$}}^{#1\atop #2}_{#3}}
\newcommand{\sln}[1]{\mathfrak{sl}(#1)}

\newcommand{\dv}[1]{\frac{\rmd}{\rmd t}}
\newcommand{\tfrac}[2]{\scriptstyle\frac{#1}{#2}\displaystyle}
\newcommand{\tensorp}{\otimes}
\newcommand{\set}[1]{\{#1\}}

\newcommand{\Complex}{\mathbb{C}}
\newcommand{\adj}[1]{#1^{\dagger}}
\newcommand{\Hilbert}{\mathcal{H}}
\newcommand{\Liouville}{\mathcal{L}}
\newcommand{\StateOp}{\hat{\rho}}
\newcommand{\Hop}[1]{\hat{#1}}
\newcommand{\Lop}[1]{\breve{#1}}
\newcommand{\Lket}[1]{\ensuremath{\bigl| #1 \bigr)}}
\newcommand{\Lbra}[1]{\ensuremath{\bigl( #1 \bigr|}}
\newcommand{\ket}[1]{\ensuremath{\bigl| #1 \bigr\rangle}}

\newcommand{\dyad}[2]{\ensuremath{\bigl| #1 \bigr\rangle\bigl\langle
    #2 \bigl|}}

\begin{document}


\title[Algebraic solution of the Lindblad equation for a collection of
 multilevel systems]{Algebraic solution of the Lindblad equation for a collection of
 multilevel systems coupled to independent environments}
 \author{Marduk Bolaños and Pablo Barberis-Blostein}
 \address{Instituto de Investigaciones en Matemáticas Aplicadas y en
 Sistemas,\\ Universidad Nacional Autónoma de México,\\ Ciudad
 Universitaria, Coyoacan 04510, Mexico City, Mexico.}
\ead{marduk@ciencias.unam.mx}



\begin{abstract}
  We consider the Lindblad equation for a collection of multilevel
  systems coupled to independent environments. The equation is
  symmetric under the exchange of the labels associated with each
  system and thus the open-system dynamics takes place in the
  permutation-symmetric subspace of the operator space. The dimension
  of this space grows polynomially with the number of systems. We
  construct a basis of this space and a set of superoperators whose
  action on this basis is easily specified. For a given number of
  levels, $M$, these superoperators are written in terms of a bosonic
  realization of the generators of the Lie algebra $\sln{M^2}$. In
  some cases, these results enable finding an analytic solution of the
  master equation using known Lie-algebraic methods. To demonstrate
  this, we obtain an analytic expression for the state operator of a
  collection of three-level atoms coupled to independent radiation
  baths. When analytic solutions are difficult to find, the basis and
  the superoperators can be used to considerably reduce the
  computational resources required for simulations.
\end{abstract}

\submitto{\JPA}


\section{Introduction}
\label{sec:introduction}

The theory of open quantum systems was developed in order to account
for dissipation and loss of coherence in quantum systems. This
framework considers a quantum system as composed of two interacting
parts: a small, experimentally accessible \emph{open quantum system}
and a large \emph{environment}. Many realistic situations are
accurately described considering that the environment has short
correlation times \cite{breuer2007theory}. The resulting Markovian
dynamics is modelled with the \emph{Lindblad master equation}
\cite{Lindblad1976,Gorini1976}, which describes the evolution of the
reduced state operator, $\StateOp$, of the open system:
\begin{equation}
  \label{eq:Lindblad_master_equation}
  \Lop{L}(\StateOp) = \dv{t}\StateOp = -\frac{\rmi}{\hbar}[\Hop{H},
  \StateOp] + \frac12
  \sum_{\alpha} \gamma_{\alpha}(2\Hop{L}_{\alpha}\StateOp\adj{\Hop{L}}_{\alpha}
  - \adj{\Hop{L}}_{\alpha}\Hop{L}_{\alpha}\StateOp
  - \StateOp\adj{\Hop{L}}_{\alpha}\Hop{L}_{\alpha}).
\end{equation}
In (\ref{eq:Lindblad_master_equation}) $\Hop{H}$ is the effective
Hamiltonian of the system, which differs from the free evolution due
to the coupling to the environment, and the Lindblad ``jump''
operators $\Hop{L}_{\alpha}$ are related to decoherence and
dissipation processes that occur with rate $\gamma_{\alpha}$. The
dagger ($\dagger$) denotes Hermitian conjugation. Since the
\emph{Liouvillian} $\Lop{L}$ acts on an operator, it is called a
\emph{superoperator}. Throughout this article, we will denote
superoperators with a breve ($\breve{A}$).

In many cases finding an analytic solution of
(\ref{eq:Lindblad_master_equation}) is not feasible and one must
resort to a numerical solution. In general, the simulation of the
Lindblad equation is computationally expensive. However, there are
some ways to reduce this cost. The quantum Monte Carlo wavefunction
method \cite{Dalibard1992,Dum1992,Tian1992} maps the problem of
simulating (\ref{eq:Lindblad_master_equation}) to a Monte Carlo
calculation of quantum trajectories in state space, which is amenable
to parallel computation. Recently a method was introduced
\cite{le_bris_low-rank_2013} that consists in approximating the
evolution of an $n \times n$ density matrix restricting the dynamics
to a set of density matrices of rank $m < n$. For many-body systems
with short-range interactions, it has been shown that the
computational cost of simulating the evolution of observables with a
finite spatial support is independent of the system size
\cite{barthel_quasilocality_2012}. The fact that a symmetric Lindblad
equation preserves the symmetry of the state operator can be exploited
to considerably reduce the computational cost of numerically solving
the equation. It has long been known \cite{Sarkar1987b} that the
symmetric state operator of a collection of two-level systems has only
$(N+1)(N+2)(N+3)/6$ independent coefficients. This result was
rediscovered in \cite{Hartmann2012} and has been exploited in recent
research \cite{Xu2013,WoodCory2015,Optimal2014,Spaser2015}. Here, it
will be shown that the symmetric state operator of $N$ $M$-level
systems is characterized by $\mathrm{poly}(N)$ parameters, thus
allowing an efficient simulation of the dynamics.

In recent years, several methods have been developed in order to
obtain an analytic solution of (\ref{eq:Lindblad_master_equation}).
These include diagonalization of the Liouvillian
\cite{Briegel1993,torres2014,li_perturbative_2014}, Lie-algebraic
methods
\cite{Wang2003,klimov_algebraic_2003,arevalo-aguilar_solution_1998,rau_embedding_2002},
extending the open system with an auxiliary system \cite{YiYu2001},
decomposing the reduced state operator in a sum of diagonal block
operators \cite{Napoli2005} and series expansions
\cite{lucas_adaptive_2013}. However, exact solutions for a collection
of systems are scarce. Recently \cite{Hartmann2012}, Hartmann obtained
an analytic expression for the state operator of a collection of
two-level atoms coupled to independent radiation baths. This involved
constructing a basis for the symmetric subspace of the operator space
and writing the master equation in terms of a set of superoperators
whose action on the above basis vectors can be easily calculated.

The present work extends the results of \cite{Hartmann2012} for a
collection of $N$ $M$-level systems. First, we build a basis for the
symmetric subspace of the operator space and show that its dimension
grows polynomially with $N$.
Then, we write the Lindblad equation in terms of a bosonic realization
of the generators of $\sln{M^2}$. The action of such superoperators on
the basis vectors found above is easily calculated. These results can
be used to find an analytic solution of the Lindblad equation using
known Lie-algebraic methods. As an example, we obtain an exact
analytic expression for the state operator of a collection of
three-level atoms coupled to independent radiation baths.

The outline of the article is as follows. In section
\ref{sec:superoperators_Liouville} we present the basics of
superoperators and their matrix representation. In section
\ref{sec:analytical_solution_symmetric_Lindblad} we briefly review the
Lie-algebraic method that will be used for solving the Lindblad master
equation. In section \ref{sec:symmetric_subspace_Liouville} we build a
basis for the symmetric subspace of the operator space and show that its
dimension grows polynomially with $N$. In section
\ref{sec:slocal_symmetric_Liouvillians} we consider the Liouvillian
for a collection of $N$ $M$-level systems subject to independent
dissipative processes and we show that it can be written in terms of a
bosonic realization of the generators of $\sln{M^2}$. In section
\ref{sec:example} we find an analytic expression for the state
operator of a collection of three-level atoms coupled to independent
radiation baths and point out some further applications of the results
of this work.



\section{Liouville space and superoperators}
\label{sec:superoperators_Liouville}

Let the Hilbert space associated with a single $M$-level system be
$\mathcal{H}_{M}$. Orthonormal basis vectors are given, in Dirac's
notation, by $\ket{1}, \ket{2}, \ldots, \ket{M}.$ The state operator
$\StateOp$ of the system, as well as any operator $\Hop{A}$ acting on
$\mathcal{H}_{M}$, are elements of a larger Hilbert space
$\mathcal{L}_{M^2} = \mathcal{H}_{M} \tensorp \mathcal{H}^{*}_{M}$ of
dimension $M^2$ called \emph{Liouville space} or \emph{von Neumann
  space} \cite{VasilyTarasov, ShaulMukamel, jeener1982superoperators}.
The asterisk denotes the dual space and $\tensorp$ is the tensor
product of vector spaces. For convenience, we introduce a Dirac-like
notation to denote ket vectors as $|\Hop{A}) \in \mathcal{L}_{M^2}$
and bra vectors as $(\Hop{A}| \in \mathcal{L}^{*}_{M^2}$.
$\mathcal{L}_{M^2}$ is equipped with the Hilbert--Schmidt scalar
product $(\Hop{A} | \Hop{B}) = \Tr(\adj{\Hop{A}}\Hop{B})$, which
determines the Hilbert--Schmidt norm
$||\Hop{A}||_2 = \sqrt{(\Hop{A} | \Hop{A})}$. The basis vectors of
$\mathcal{H}_{M}$ induce an orthonormal basis for $\mathcal{L}_{M^2}$
given by the ket-bra operators $\dyad{m}{n}$ with
$m,n \in (1,2,\ldots,M)$, which we denote as
$|mn) := \left|\dyad{m}{n}\right)$. In terms of this basis we define
two matrix representations of an element of $\mathcal{L}_{M^2}$ given
by:
\begin{eqnarray}
  \label{eq:matrix_rep_Liouville_ket}
  |\Hop{A}) = \sum_{m,n = 1}^{M} (mn|\Hop{A})|mn), \quad & A_{mn} &:= (mn|\Hop{A}),\nonumber\\
  |\Hop{A}) = \sum_{\alpha=1}^{M^2} (\alpha|\Hop{A})|\alpha),
              \quad & A_{\alpha} &:= (\alpha|\Hop{A}).
\end{eqnarray}
The relationship between the matrix elements $A_{mn}$ and $A_{\alpha}$
depends on the choice of linear map transforming an $M \times M$
matrix $\bi{A}$ into an $M^2 \times 1$ vector. Two such commonly used
maps are given by \cite{horn1994topics}
\begin{eqnarray}
  \label{eq:vec_vecT}
  \fl
  \mathrm{col}\bi{A} &:= \mathrm{vec}\bi{A} &=
  [A_{11},\ldots,A_{M1},A_{12},\ldots,A_{M2},\ldots,A_{1M},\ldots,A_{MM}]^T,\nonumber\\
  \fl
  \mathrm{row}\bi{A} &:= \mathrm{vec}\bi{A}^T &=
  [A_{11},\ldots,A_{1M},A_{21},\ldots,A_{2M},\ldots,A_{M1},\ldots,A_{MM}]^T,
\end{eqnarray}
where the former stacks the columns of $\bi{A}$ from left to right and
the latter stacks the rows of $\bi{A}$ from top to bottom. In order to
illustrate (\ref{eq:matrix_rep_Liouville_ket}) and (\ref{eq:vec_vecT})
we consider a $2 \times 2$ matrix and its row representation:
\begin{equation}
  \label{eq:example_row_A}
  \bi{A} = \left(
    \begin{array}{cc}
      A_{22} & A_{21}\\[1mm]
      A_{12} & A_{11}
    \end{array}
  \right)
 \overset{\mathrm{row}}{\longrightarrow} \left[
    \begin{array}{c}
      \Lbra{22}\Hop{A})\\[1mm]
      \Lbra{21}\Hop{A})\\[1mm]
      \Lbra{12}\Hop{A})\\[1mm]
      \Lbra{11}\Hop{A})
    \end{array}
  \right]
 = \left[
    \begin{array}{c}
      \Lbra{1}\Hop{A})\\[1mm]
      \Lbra{2}\Hop{A})\\[1mm]
      \Lbra{3}\Hop{A})\\[1mm]
      \Lbra{4}\Hop{A})
    \end{array}
  \right].
\end{equation}

Linear maps $\Lop{T}: \mathcal{L}_{M^2} \rightarrow \mathcal{L}_{M^2}$
are usually called superoperators and represent the transformation
$\Lop{T}|\Hop{A}) = |\Hop{B})$. Their matrix representation as an
$M^2 \times M^2$ matrix $\bi{T} = [T_{\alpha\beta}]$ is defined as:
\begin{equation}
  \label{eq:matrix_rep_superoperator}
  \Lop{T} = \sum_{\alpha, \beta = 1}^{M^2}
  (\alpha|\Lop{T}|\beta)|\alpha)(\beta|, \quad T_{\alpha\beta}:= (\alpha|\Lop{T}|\beta),
\end{equation}
where $|\alpha)(\beta|$ is an orthonormal basis in the space of
superoperators, induced by the basis in Liouville space. For
consistency $\bi{T}$ will be called a supermatrix. Every
superoperator may be written as \cite{ernst1990principles}
\begin{equation}
  \label{eq:general_superoperator}
  \Lop{T} = \sum_{ij}t_{ij}\Lop{V}_i^{L}\Lop{V}_j^{R},
\end{equation}
where $\Lop{V}_i^{L}\Lket{\Hop{A}} = \Lket{\Hop{V}_i\Hop{A}}$,
$\Lop{V}_j^{R}\Lket{\Hop{A}} = \Lket{\Hop{A}\Hop{V}_j}$ and
$\set{\Hop{V}_r, r = 1,2,\ldots,M^2}$ is a complete set of basis
operators. Therefore, a general operator transformation is of the form
\begin{equation}
  \label{eq:general_operator_transformation}
  \Lop{T}\Lket{\Hop{A}} = \sum_{ij}t_{ij} |\Hop{V}_i\Hop{A}\Hop{V}_j).
\end{equation}

The linear maps in (\ref{eq:vec_vecT}) induce two matrix
representations of $\bi{V}_i\bi{A}\bi{V}_j$
\cite{havel2003robust}:
\begin{eqnarray}
  \label{eq:matrix_rep_MAN}
  \mathrm{col}\;\;\bi{V}_i\bi{A}\bi{V}_j &=
  (\bi{V}_j^T\tensorp\bi{V}_i)\mathrm{col}\;\bi{A}, \quad \bi{T}_{\mathrm{col}} &=
  \bi{V}_j^T\tensorp\bi{V}_i,\nonumber\\
  \mathrm{row}\bi{V}_i\bi{A}\bi{V}_j &=
  (\bi{V}_i\tensorp\bi{V}_j^T)\mathrm{row}\bi{A}, \quad \bi{T}_{\mathrm{row}} &=
  \bi{V}_i\;\tensorp\bi{V}_j^T.
\end{eqnarray}

We consider now the Hilbert space associated with a collection of
$N$ $M$-level systems $\Hilbert^{\tensorp N}_{M} := \Hilbert^{(1)}_M
\tensorp \Hilbert^{(2)}_M \tensorp \cdots \tensorp \Hilbert^{(N)}_M$.
The basis of $\Hilbert_M$ induces a basis of this space given by the
ordered set $\{\ket{h_1}\ket{h_2}\cdots\ket{h_N}, h_i = 1,2,\ldots,
M\}$. The dimension of $\Hilbert^{\tensorp N}_{M}$ is therefore $M^N$.
The associated Liouville space is defined as $\Liouville^{\tensorp
  N}_{M^{2}} := \Liouville_{M^2}^{(1)} \tensorp \Liouville_{M^2}^{(2)}
\tensorp \cdots \tensorp \Liouville_{M^2}^{(N)}$. The basis of
$\Liouville_{M^2}$ induces a basis of this space given by the ordered
set
$\{\Lket{l_1l_1'}\Lket{l_2l_2'}\cdots\Lket{l_Nl_N'},
l_i, l_i' = 1,2,\ldots M\}$. The dimension of
$\Liouville^{\tensorp N}_{M^{2}}$ is therefore $M^{2N}$.

\section{Lie-algebraic solution of the Lindblad equation}
\label{sec:analytical_solution_symmetric_Lindblad}

In this section we describe a known algebraic method for finding an
analytic solution to the initial value problem
\begin{equation}
  \label{eq:master_equation_sloc}
  \dv{t}\StateOp = \Lop{L}\StateOp, \quad \StateOp(0) = \StateOp_0,
\end{equation}
where $\Lop{L}$ is a time-independent linear combination of the
generators of a finite-dimensional Lie algebra. Therefore, the formal
solution to (\ref{eq:master_equation_sloc}) is given by
$\StateOp(t) = \exp(\Lop{L} t)\StateOp_0$.

The exponential $e^{tA}$ of an element $A$ of a finite-dimensional Lie
algebra spanned by a set of $n$ generators $\set{L_j}$, yields an
element $\mathcal{A}$ of the corresponding Lie group, which is
parametrized in terms of \emph{canonical coordinates of the first
  kind} \cite{varadarajan2013lie}:
\begin{equation}
  \label{eq:canonical_coords_1st_kind}
  \mathcal{A}_{1}(\balpha) = \exp \left[t\sum_{j=1}^n \alpha_j L_j\right].
\end{equation}
Another common parametrization involves \emph{canonical coordinates of
  the second kind}:
\begin{equation}
  \label{eq:canonical_coords_2nd_kind}
  \mathcal{A}_{2}(\bbeta) = \prod_{j=1}^n\exp(\beta_j(t) L_j).
\end{equation}
The parameters $\balpha$ and $\bbeta(t)$ are related through analytic
expressions called Baker-Campbell-Hausdorff (BCH) formulas
\cite{gilmore2012lie}. These formulas are usually obtained by solving
a set of $n$ coupled, nonlinear, ordinary differential equations,
called Wei-Norman equations \cite{Wei1963}. Recently, it was shown
\cite{charzynski_weinorman_2013} that for the Lie algebra
$\mathfrak{sl}(n,\Complex)$ 
this nonlinear system can be reduced to a hierarchy of matrix Riccati
equations. While it is known that matrix Riccati equations of the
projective type with constant coefficients may be readily integrated
\cite{bountis_integrability_1986}, one must bear in mind that there is
no general method for obtaining the solution of a Riccati equation
with time-dependent coefficients. However, approximate solutions can
be obtained using the Taylor matrix method \cite{gulsu_solution_2006}
and the variational iteration method \cite{batiha_application_2007}.
Moreover, the number of Wei-Norman equations that have to be
integrated may be reduced \cite{echave_baker-campbell-hausdorff_1992}.

Obtaining BCH formulas is simple in certain cases. It is known that
they can always be obtained for solvable Lie algebras
\cite{wei_global_1964,JurgenFuchs}. Moreover, for small
dimensions Riccati equations can be avoided, since calculation
\cite{cheng_more_1997} of the matrix exponentials in
(\ref{eq:canonical_coords_1st_kind}) and
(\ref{eq:canonical_coords_2nd_kind}) yields a system of algebraic
nonlinear equations, that can be solved for the coefficients
$\bbeta(t)$. In section \ref{sec:example} we will make use of this
method. We remark that since the product of exponentials
(\ref{eq:canonical_coords_2nd_kind}) is --in general-- not global
\cite{wei_global_1964}, care must be taken with the domain of validity
of the BCH formulas obtained by any method.

\section{The symmetric subspace of Liouville space}
\label{sec:symmetric_subspace_Liouville}

We call a state operator that is invariant under the exchange of the
labels associated with individual systems a \emph{symmetric state
  operator} and denote it $|\StateOp_{\mathrm{sym}})$. Since symmetric
state operators are elements of
$\mathcal{S}(\Liouville_{M^2}^{\tensorp N})$, the symmetric subspace
of the Liouville space, the action of a symmetric Liouvillian
$\Lop{L}_{\mathrm{sym}}$ on $|\StateOp_{\mathrm{sym}})$ will result in
another element of $\mathcal{S}(\Liouville_{M^2}^{\tensorp N})$. That
is,
$\Lop{L}_{\mathrm{sym}}: \mathcal{S}(\Liouville_{M^2}^{\tensorp N})
\rightarrow \mathcal{S}(\Liouville_{M^2}^{\tensorp N})$.

A basis vector of $\mathcal{S}(\Liouville_{M^2}^{\tensorp N})$ is
given by
\begin{equation}
  \label{eq:symmetric_vector}
  |S)_{\set{n_{ij}}} = \mathrm{K} \sum_P \Lop{P} \bigotimes_{i,j=1}^M |ij)^{\tensorp
    n_{ij}}, \quad \mathrm{K} = \frac{1}{N!} \prod_{i,j=1}^M n_{ij}!,
\end{equation}
where $|ij)$ are unit basis vectors in $\mathcal{L}_{M^2}$,
$n_{ij} = 0,1,\ldots,N$, $\sum_{i,j=1}^M n_{ij} = N$ and for nonzero
$n_{ij}$, $|ij)^{\tensorp n_{ij}} = \bigotimes_{k=1}^{n_{ij}} |ij)_k$.
In (\ref{eq:symmetric_vector}) the index $P$ runs over all
permutations of two vectors $|ij), |i'j')$ that yield distinct tensor
products and $\Lop{P}$ is a superoperator that performs a given
permutation. The constant $K$ accounts for the permutations that yield
repeated tensor products and ensures that the symmetric basis vectors
have unit norm.

For three-level ($\ket{0}, \ket{1}, \ket{2}$) systems we introduce the
following notation for symmetric basis vectors
\begin{equation}
  \label{eq:Q_notation}
  |S)_{\set{n_{ij}}} := \raisebox{-1pt}{\Large{Q}}\arraycolsep=1.5pt%
   \scriptsize{\begin{array}{lll}
      n_{00} & n_{01} & \\
      n_{10} & n_{11} & n_{12}\\
      n_{20} & n_{21} & n_{22}\\
  \end{array}}\normalsize
\end{equation}
that will be convenient in section \ref{sec:example}. In
(\ref{eq:Q_notation}), we have arbitrarily chosen $n_{02}$ to depend
on both the other indices and $N$ and have thus omitted it. To
illustrate (\ref{eq:Q_notation}) we consider $N = 3$, $n_{01} = 2$ and
$n_{22} = 1$:
\begin{equation}
  \label{eq:example_symmetric_basis_vector}
  \fl  \raisebox{-1pt}{\Large{Q}}\arraycolsep=1.5pt%
  \scriptsize{\begin{array}{lll}
      0 & 2 & \\
      0 & 0 & 0\\
      0 & 0 & 1\\
    \end{array}}\normalsize
  = \frac13 \left[|22)_1|01)_2|01)_3 + |01)_1|01)_2|22)_3 + |01)_1|22)_2|01)_3\right].
\end{equation}
We remark that some of the symmetric basis vectors are devoid of
physical meaning, since they do not satisfy the unit trace condition.

The number of sets $\set{n_{ij}}$ that characterize all possible
symmetric basis vectors is the dimension of
$\mathcal{S}(\mathcal{L}_{M^2}^{\tensorp N})$. It may be calculated as
the number of ways to distribute $N$ objects in $M^2$ bins,
considering that any bin can be empty:
\begin{equation}
  \label{eq:dimension_symmetric_subpsace}
  s := {N + M^2 - 1 \choose N} = \frac{1}{(M^2 - 1)!} \prod_{k =
    0}^{M^2-2} (N+M^2-1-k).
\end{equation}
Therefore, grouping the basis vectors of the Liouville space
$\mathcal{L}_{M^2}^{\tensorp N}$ into symmetric linear combinations
results in a subspace whose dimension is polynomial in $N$.

\begin{figure}[t]
 \centering\includegraphics[scale=0.5]{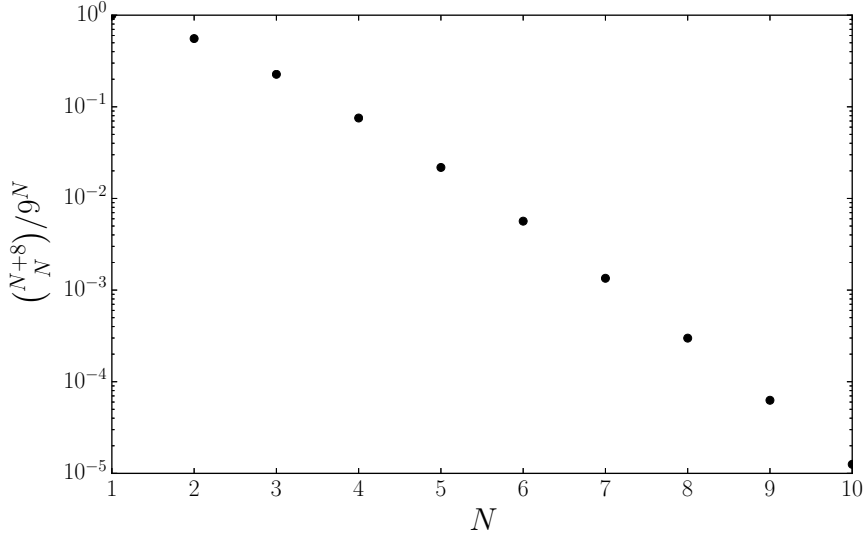}
  \caption{Comparison between the dimensions of the Liouville space
    and its symmetric subspace for $N$ three-level systems.}
  \label{fig:comparison_of_dimensions}
\end{figure}

In terms of symmetric basis vectors, a symmetric state operator is
expressed as
\begin{equation}
  \label{eq:symmetric_state_operator}
  |\StateOp_{\mathrm{sym}}) = \sum_{k=1}^s c_k |S_{n_k}),
\end{equation}
where the $c_k$ must be such that $\StateOp_{\mathrm{sym}}$ has the
properties of a state operator and $n_k$ denotes one of the $s$ sets
$\set{n_{ij}}$. From~(\ref{eq:dimension_symmetric_subpsace}) it is
clear that $|\StateOp_{\mathrm{sym}})$ has only $O(N^{M^2-1})$
independent parameters and therefore the computational resources
required for the simulation of the symmetric Lindblad equation are
substantially reduced compared to a simulation in the whole Liouville
space (see Figure \ref{fig:comparison_of_dimensions}).

\section{Symmetric Liouvillians with independent dissipative processes}
\label{sec:slocal_symmetric_Liouvillians}

In the following we shall be concerned with the algebraic structure of
symmetric Liouvillians with independent dissipative processes (SInDiP
Liouvillians) defined as:
\begin{equation}
  \label{eq:strictly_local_Liouvillian}
  \Lop{L}(\StateOp) = \sum_{\mu=1}^N \Lop{L}^{\mu}(\StateOp) =  -\frac{i}{\hbar}\sum_{\mu=1}^N\Lop{H}^{\mu}
  (\StateOp) + \sum_{\mu=1}^N\Lop{D}^{\mu}(\StateOp),
\end{equation}
with the superoperators
\begin{equation}
  \label{eq:Hamiltonian}
   \Lop{H}^{\mu}(\StateOp) = \left[\sum_{k=1}^{M^2-1}
    h_k\Hop{F}^{\mu}_k, \StateOp\right] = \sum_{k=1}^{M^2-1}
    h_k \left( [\Hop{F}^{\mu}_k]^L\mathds{1}^R - \mathds{1}^L[\Hop{F}^{\mu}_k]^R \right)\StateOp
\end{equation}
and
\begin{eqnarray}
  \label{eq:dissipator}
   \Lop{D}^{\mu}(\StateOp) &= \frac12 \sum_{j,k=1}^{M^2-1} a_{jk} 
  (2\Hop{F}_j^{\mu}\StateOp \Hop{F}_{\,\,\,k}^{\dagger\mu} -
  \Hop{F}_{\,\,\,k}^{\dagger\mu}\Hop{F}_j^{\mu}\StateOp -
  \StateOp\Hop{F}_{\,\,\,k}^{\dagger\mu}\Hop{F}_j^{\mu})\\
   &= \frac12 \sum_{j,k=1}^{M^2-1} a_{jk} 
  \left( 2[\Hop{F}_j^{\mu}]^L [\Hop{F}_{\,\,\,k}^{\dagger\mu}]^R -
  [\Hop{F}_{\,\,\,k}^{\dagger\mu}\Hop{F}_j^{\mu}]^L \mathds{1}^R -
  \mathds{1}^L [\Hop{F}_{\,\,\,k}^{\dagger\mu}\Hop{F}_j^{\mu}]^R
        \right)\StateOp, \nonumber
\end{eqnarray}
where we used one of the standard forms of the dissipator $\Lop{D}$,
which is equivalent to the diagonal form used in
(\ref{eq:Lindblad_master_equation}). The coefficients $h_k$ in
(\ref{eq:Hamiltonian}) have to ensure the Hermiticity of the
Hamiltonian and the coefficients $a_{jk}$ in (\ref{eq:dissipator}) are
the entries of a complex, Hermitian, positive-semidefinite matrix. The
operators $\Hop{F}_j, j=1,2,\ldots,M^2-1$ are traceless, orthonormal
(with respect to the Hilbert-Schmidt inner product) and form a
complete set. Since they are not necessarily Hermitian, they can be
regarded as elements of the operator realization of the complex Lie
algebra $\sln{M}$ of traceless $M \times M$ matrices. The properties
of the operators $\Hop{F}_j$ ensure that the decomposition of the
generator $\Lop{L}$ of a Markovian master equation into a Hamiltonian
and a dissipator is unique \cite{Gorini1976}.

Following \cite{Ticozzi28112012} we use the term \emph{strictly local
  operators} for the elements of the set
\begin{equation}
  \label{eq:strictly_local_operators}
  \fl\zeta = \left\{ \mathds{1}^1 \tensorp \mathds{1}^2 \tensorp \cdots \tensorp
    \Hop{O}^{\mu}_k \tensorp \cdots \tensorp \mathds{1}^N, \mu =
    1,2,\ldots,N; k = 1,2,\ldots,M^2-1 \right\},
\end{equation}
where $\Hop{O}^{\mu}_k$ acts on the $\mu$-th
system. The operator transformation $\Lop{L}(\StateOp)$ is a sum of
$N$ strictly local terms $\Lop{L}^{\mu}(\StateOp)$. Therefore we may
write (\ref{eq:strictly_local_Liouvillian}) in terms of the
\emph{collective superoperators} ($q = 1,2,\ldots, (n^2 - n)/2$ and $p = 1,2,\ldots, n-1$)
\begin{equation}
  \label{eq:collective_superoperators}
  \Lop{A}^{q}_{\pm} := \sum_{\mu=1}^N (\mathbb{A}^{q}_{\pm})^{\mu},\quad
  \Lop{A}^{p}_{3} := \sum_{\mu=1}^N(\mathbb{A}^{p}_{3})^{\mu},
\end{equation}
where 
\begin{eqnarray}
  \label{eq:collective_superops_A}
  (\mathbb{A}^{q}_{\pm})^{\mu} = \underbrace{\mathds{1} \tensorp
    \mathds{1} \tensorp \cdots \tensorp
    \mathds{1}}_{\mu-1} \tensorp \mathbb{A}^{q}_{\pm} \tensorp
  \underbrace{\mathds{1} \tensorp \cdots \tensorp \mathds{1}}_{N-\mu},\\
  (\mathbb{A}^{p}_{3})^{\mu} = \underbrace{\mathds{1} \tensorp
    \mathds{1} \tensorp \cdots \tensorp
    \mathds{1}}_{\mu-1} \tensorp \mathbb{A}^{p}_{3} \tensorp
  \underbrace{\mathds{1} \tensorp \cdots \tensorp \mathds{1}}_{N-\mu}
\end{eqnarray}
and the superoperators $\mathbb{A}^{q}_{\pm}$ and $\mathbb{A}^{p}_{3}$
are a realization of the generators of $\sln{n = M^2}$.

In order to find an analytic expression for the symmetric state
operator at time $t$, it will be necessary to calculate the action of
the Liouvillian on the symmetric basis vectors
(\ref{eq:symmetric_vector}). We recall from the theory of angular
momentum that it is easier to calculate the matrix elements of
operators acting on the symmetric representation of $\mathfrak{u}(n)$
if the operators are written in terms of the Jordan-Schwinger boson
representation. The bosonization of a generator $\bi{O}_{\alpha}$ of a
Lie algebra of dimension $n$ is accomplished with the map
\cite{biedenharn2009angular}:
\begin{equation}
  \label{eq:bosonization}
\mathcal{B}(\bi{O}_{\alpha}):=   \sum_{j,k = 1}^{n}
  \Hop{b}^{\dagger}_j (\bi{O}_{\alpha})_{jk} \Hop{b}_{k} \, ,
\end{equation}
where the operators $\Hop{b}$ satisfy the usual bosonic commutation
relations $[\Hop{b}_i,\Hop{b}^{\dagger}_j] = \delta_{ij}$ and
$[\Hop{b}_i,\Hop{b}_j] = [\Hop{b}^{\dagger}_i,\Hop{b}^{\dagger}_j] =
0$,
and the products $\Hop{b}_i^{\dagger}\Hop{b}_j$ satisfy the
commutation relations of the generators of the algebra
$\mathfrak{u}(n)$:
\begin{equation}
  \label{eq:bilinear_commutator}
  [\Hop{b}_i^{\dagger}\Hop{b}_j,\Hop{b}_m^{\dagger}\Hop{b}_n] =
  \Hop{b}_i^{\dagger}\Hop{b}_n \delta_{jm} -
  \Hop{b}_m^{\dagger}\Hop{b}_j \delta_{in}.
\end{equation}
The map (\ref{eq:bosonization}) has the property of preserving all
commutators,
$\mathcal{B}([\bi{O}_{\alpha}, \bi{O}_{\beta}]) =
[\mathcal{B}(\bi{O}_{\alpha}), \mathcal{B}(\bi{O}_{\beta})]$.
Therefore, it is a Lie algebra homomorphism. We remark that this
procedure is only introduced as an algebraic device. The boson
operators should not be understood in the sense of quantum field
theory. We also note that in the case of Liouville space, the boson
operators are actually superoperators acting on the unit basis vectors
$|ij)$ defined in section \ref{sec:symmetric_subspace_Liouville}.

In order to better understand the bosonization procedure in the
present context, we consider the superoperator
$\Hop{\sigma}_{20,+}^L\mathds{1}^R$ acting on a three-level system.
The nonzero matrix elements of this superoperator read:
\begin{equation}
  \label{eq:nonzero_matrix_elements}
  (20|\Hop{\sigma}_{20,+}^L\mathds{1}^R|00) =
  (21|\Hop{\sigma}_{20,+}^L\mathds{1}^R|01) =
  (22|\Hop{\sigma}_{20,+}^L\mathds{1}^R|02) = 1,
\end{equation}
where we used (\ref{eq:general_operator_transformation}) to calculate
$\Hop{U}^L\Hop{V}^R|ij) = |\Hop{U}\dyad{i}{j}\Hop{V}) = |i'j')$.
Therefore, the bosonization mapping yields
\begin{equation}
  \label{eq:bosonization_example}
  \Hop{\sigma}_{20,+}^L\mathds{1}^R \mapsto
  \adj{\Lop{b}}_{20}\Lop{b}_{00} + \adj{\Lop{b}}_{21}\Lop{b}_{01} + \adj{\Lop{b}}_{22}\Lop{b}_{02}.
\end{equation}

In terms of bosonic superoperators, the collective superoperators
defined in (\ref{eq:collective_superoperators}) have the form
($\ell_i > \ell_j$)
\begin{equation}
  \label{eq:bosonic_superoperators}
  \fl \Bop{\ell_i}{\ell_j}{+} = \sum_{\mu=1}^N \Lop{b}_{\ell_i}^{\dagger(\mu)}\Lop{b}_{\ell_j}^{(\mu)}, \quad \Bop{\ell_i}{\ell_j}{-} =
  \sum_{\mu=1}^N \Lop{b}_{\ell_j}^{\dagger(\mu)}\Lop{b}_{\ell_i}^{(\mu)}, \quad \Bop{\ell_i}{\ell_j}{0} =
  \sum_{\mu=1}^N \frac12\left(\Lop{b}_{\ell_i}^{\dagger(\mu)}\Lop{b}_{\ell_i}^{(\mu)} - \Lop{b}_{\ell_j}^{\dagger(\mu)}\Lop{b}_{\ell_j}^{(\mu)}\right),
\end{equation}
where $\ell_i$ and $\ell_j$ denote the labels of two unit basis
vectors $|ii')$ and $|jj')$. We remark that the elements of the set
$\set{\Bop{\ell_i}{\ell_j}{0}}$ are not linearly independent and,
therefore, it is necessary to choose a subset thereof, denoted as
$\set{\Bop{\ell_i}{\ell_j}{3}}$, comprising $M^2-1$ superoperators.
Moreover, using (\ref{eq:bilinear_commutator}) it follows that the
collective superoperators (\ref{eq:bosonic_superoperators}) form
$\sln{2}$ subalgebras of $\sln{M^2}$:
\begin{equation}
  \label{eq:sl2_commutators}
  \Bop{\ell_i}{\ell_j}{0} := \frac12 [\Bop{\ell_i}{\ell_j}{+}, \Bop{\ell_i}{\ell_j}{-}], \quad [\Bop{\ell_i}{\ell_j}{0}, \Bop{\ell_i}{\ell_j}{\pm}] = \pm\Bop{\ell_i}{\ell_j}{\pm}.
\end{equation}
This realization of the generators of $\sln{M^2}$ simplifies the
calculation of the action of a collective superoperator on a symmetric
vector. For example, using the notation introduced in
(\ref{eq:Q_notation}) for three-level systems:
\begin{eqnarray}
\label{eq:superoperators_on_basis_vectors}
\Bop{22}{21}{+}\raisebox{-1pt}{\Large{Q}}\arraycolsep=1.5pt%
  \scriptsize{\begin{array}{lll}
      n_{00} & n_{01} & \\
      n_{10} & n_{11} & n_{12}\\
      n_{20} & n_{21} & n_{22}\\
    \end{array}}\normalsize = n_{21}\normalsize\raisebox{-1pt}{\Large{Q}}\arraycolsep=1.5pt%
      \scriptsize{\begin{array}{lll}
      n_{00} & n_{01} & \\
      n_{10} & n_{11} & n_{12}\\
      n_{20} & n_{21}-1 & n_{22}+1\\
        \end{array}}\normalsize, \nonumber\\
\Bop{22}{21}{-}\raisebox{-1pt}{\Large{Q}}\arraycolsep=1.5pt%
  \scriptsize{\begin{array}{llll}
      n_{00} & n_{01} & \\
      n_{10} & n_{11} & n_{12}\\
      n_{20} & n_{21} & n_{22}\\
    \end{array}}\normalsize = n_{22}\normalsize\raisebox{-1pt}{\Large{Q}}\arraycolsep=1.5pt%
      \scriptsize{\begin{array}{llll}
      n_{00} & n_{01} & \\
      n_{10} & n_{11} & n_{12}\\
      n_{20} & n_{21}+1 & n_{22}-1\\
        \end{array}}\normalsize, \\
\Bop{22}{21}{0}\raisebox{-1pt}{\Large{Q}}\arraycolsep=1.5pt%
  \scriptsize{\begin{array}{llll}
      n_{00} & n_{01} & \\
      n_{10} & n_{11} & n_{12}\\
      n_{20} & n_{21} & n_{22}\\
    \end{array}}\normalsize = \frac12(n_{22} - n_{21})\raisebox{-1pt}{\Large{Q}}\arraycolsep=1.5pt%
      \scriptsize{\begin{array}{llll}
      n_{00} & n_{01} & \\
      n_{10} & n_{11} & n_{12}\\
      n_{20} & n_{21} & n_{22}\\
        \end{array}}\normalsize.\nonumber
\end{eqnarray}

In the following section we will make use of the main results of this
article, contained in this section and the previous section, in order
to obtain an analytic expression for the solution of a Lindblad
equation with a SInDiP Liouvillian. For quick reference we outline the
procedure here:
\begin{itemize}
\item Starting from a master equation in the usual notation
  (\ref{eq:strictly_local_Liouvillian}) bosonize each superoperator of
  the form $\Hop{U}^L\Hop{V}^R$ and write the master equation in terms
  of the collective superoperators in
  (\ref{eq:bosonic_superoperators}).
\item Write the solution of the master equation as a product of, in
  general entangled,  exponentials remembering that
  $e^{A + B} = e^A e^B$ iff $[A,B] = 0$. Obtain the appropriate BCH
  formulas that allow disentangling the exponentials.
\item Calculate the action of the superoperator $e^{\Lop{L}t}$ on a
  symmetric basis vector.
\end{itemize}


\section{Example: three-level atoms}
\label{sec:example}

In this section we will use the results of the preceding sections to
calculate an analytic expression for the symmetric state operator of a
collection of $N$ three-level atoms in the $\Lambda$ configuration
with energy levels $E_0 < E_1 < E_2$, interacting with a radiation
bath. Assuming orthogonal dipole moments and neglecting the
environment-mediated coupling between atoms, the corresponding
Lindblad master equation is given by \cite{Wolf2010}
\begin{equation}
  \label{eq:master_eq_lambda}
  \dv{t}\StateOp = \Lop{L}(\StateOp) =
  \sum_{\mu=1}^N\Lop{L}^{\mu}(\StateOp) = \sum_{\mu=1}^N -\frac{i}{\hbar}[\Hop{H}^{\mu},\StateOp] + (\Lop{L}_{21}^{\mu} + \Lop{L}_{20}^{\mu})\StateOp,
\end{equation}
with the Lindbladians
\begin{eqnarray}
  \label{eq:atomic_lindbladian}
  \Lop{L}_{ij}^{\mu} [\bullet] &=  (N_0+1)\gamma_{ij}\left(\Hop{\sigma}_{ij-}^{\mu}
                                 [\bullet]\Hop{\sigma}_{ij+}^{\mu}
  - \frac12\{\Hop{\sigma}_{ij+}^{\mu}\Hop{\sigma}_{ij-}^{\mu},[\bullet]\}\right)\nonumber\\
  &+
  N_0\gamma_{ij}\left(\Hop{\sigma}_{ij+}^{\mu}[\bullet]\Hop{\sigma}_{ij-}^{\mu}
  -
  \frac12\{\Hop{\sigma}_{ij-}^{\mu}\Hop{\sigma}_{ij+}^{\mu},[\bullet]\}\right),
\end{eqnarray}
where $\{A,B\} = AB + BA$ and $[\bullet]$ is a placeholder for the
operand of the superoperator. 
The Hamiltonian $\Hop{H}^{\mu}$, taking $\bar{E} = \frac13(E_0 + E_1 +
E_2)$ as the zero of energy, is
\begin{equation}
  \label{eq:three_levels_hamiltonian}
  \Hop{H}^{\mu} = \tilde{E}_{0}\Hop{\sigma}_3^{02(\mu)} +
  \tilde{E}_{1}\Hop{\sigma}_3^{12(\mu)}, \quad \tilde{E}_{i} = 2(\bar{E} - E_i).
\end{equation}
The index $\mu$ denotes that an operator acts on the Hilbert space of
the atom with label $\mu$ and we define $\Hop{\sigma}_{ij-} =
\dyad{j}{i}$ and $\Hop{\sigma}_3^{ij} = \frac12(\dyad{j}{j} -
\dyad{i}{i})$. Moreover, the products
$\Hop{\sigma}_{ij-}\Hop{\sigma}_{ij+} = \dyad{j}{j}$ may be written as
\cite{AndreiB.Klimov}:
\begin{equation}
  \label{eq:dyad_jj}
  \dyad{j}{j} \;= \frac{\mathds{1}}{3} + 2\sum_{k=0}^{j-1}\Hop{\sigma}_3^{k,k+1} - \frac{2}{3}\sum_{k=0}^{1}(2-k)\Hop{\sigma}_3^{k,k+1}.
\end{equation}

Since the operators
$\set{\Hop{\sigma}_{\pm}^{21}, \Hop{\sigma}_{\pm}^{20},
  \Hop{\sigma}_{\pm}^{10}, \Hop{\sigma}_{3}^{12},
  \Hop{\sigma}_{3}^{02}}$
are a realization of the generators of $\sln{3}$, in general, we can
construct 108 bosonized collective superoperators
$\Lop{A}^{\ell_i\ell_j}_{\pm,0}$ with
$\ell_i,\ell_j = 00,01,02,10,11,12,20,21,22$. However, as pointed out
in section \ref{sec:slocal_symmetric_Liouvillians}, from the set of 36
superoperators $\Bop{\ell_i}{\ell_j}{0}$ one must choose a subset
of eight linearly independent superoperators
$\set{\Bop{\ell_i}{\ell_j}{3}}$. The set
$\set{\Bop{\ell_i}{\ell_j}{\pm}} \cup
\set{\Bop{\ell_i}{\ell_j}{3}}$
is a realization of the generators of $\sln{9}$ and its elements are
shown in the appendix.

In terms of bosonized collective superoperators,
(\ref{eq:master_eq_lambda}) is written as
\begin{eqnarray}
  \label{eq:master_equation_superoperators}
  \fl \dv{t}\Lket{\StateOp} = \Bigl[N_{{0}} \gamma_{{20}}
  \breve{S}^{20}_{{+}} + N_{{0}} \gamma_{{21}} \breve{S}^{{21}}_{{+}}
  + \gamma_{{20}}N^{{+}}_{{0}}\breve{S}^{{20}}_{{-}} +
  \gamma_{{21}}N^{{+}}_{{0}}\breve{S}^{{21}}_{{-}} + \frac23\left(
    N^{{-}}_{{0}} \gamma_{{20}} -
    \gamma_{{21}}\tilde{N}_{{0}}\right)\breve{S}^{{20}}_{3} \nonumber\\ 
  \fl  + \frac23\left(N^{{-}}_{{0}}\gamma_{{21}} -
    \gamma_{{20}}\tilde{N}_{{0}}\right) \breve{S}^{{21}}_{3}  
  + 3\frac{i}{\hbar}\tilde{E}_0 (\Bop{12}{10}{3} -
  \Bop{21}{01}{3}) + 2\frac{i}{\hbar}
  (E_{{2}} - E_{{0}})(\Bop{02}{01}{3} - \Bop{20}{10}{3}) \nonumber\\ 
  \fl 
  + \frac13\left(N^{{-}}_{{0}} \gamma_{{20}} -
    \gamma_{{21}} \tilde{N}_{{0}}\right)
  (\Bop{12}{10}{3} + \Bop{21}{01}{3})  
  +
  \frac13\left(N^{{-}}_{{0}} \gamma_{{21}} - \gamma_{{20}}
    \tilde{N}_{{0}}\right) (\Bop{02}{01}{3} + \Bop{20}{10}{3}) \\
  \fl - 4\frac{i}{\hbar}\left(2 E_{{0}} +
    E_{{2}} - 3 \bar{E}\right)
  \Bop{21}{12}{3} -
  \frac13\gamma\tilde{N}_{{0}}N\mathds{1} \Bigr]\Lket{\StateOp} := \Lop{L}\Lket{\StateOp} \nonumber,
\end{eqnarray}
where $\gamma = \gamma_{20} + \gamma_{21}$, $\tilde{N}_0 = 2N_0 + 1$,
$N_0^{\pm} = N_0 \pm 1$, and we defined
\begin{equation}
  \label{eq:superoperators_sl3}
  \Lop{S}^{21}_{\pm,3} := \Bop{22}{11}{\pm,3}, \quad \Lop{S}^{20}_{\pm,3}
  := \Bop{22}{00}{\pm,3}, \quad \Lop{S}^{10}_{\pm,3} := \Bop{11}{00}{\pm,3}.
\end{equation}

The set consisting of eight linearly independent superoperators
\begin{equation}
  \label{eq:A3_superoperators}
  \set{\Bop{\ell_i}{\ell_j}{3}} := \set{\Bop{22}{11}{3},  
    \Bop{21}{12}{3}, \Bop{20}{10}{3}, \Bop{02}{01}{3}} \cup 
  \set{\Bop{22}{00}{3},\Bop{21}{01}{3}, \Bop{12}{10}{3}, \Bop{11}{02}{3}}
\end{equation}
was constructed considering two quartets (sets consisting of four
elements) of $\sln{2}$ subalgebras of $\sln{9}$, such that in each
quartet the generators of any algebra commute with the generators of
the other algebras. We remark that any such quartet may be used to
label the symmetric basis vectors in terms of the eigenvalues of the
Cartan and Casimir superoperators of the algebras.

Since $\Lop{L}$ is time-independent, the formal solution of
(\ref{eq:master_equation_superoperators}) is given by
\begin{eqnarray}
  \label{eq:master_equation_solution}
  \fl  \Lket{\StateOp(t)} = \e^{\alpha_1t}\exp(\alpha_2\Bop{21}{12}{3}t)\exp(\alpha_3\Bop{21}{01}{3}t)\exp(\alpha_4\Bop{20}{10}{3}t)\exp(\alpha_5\Bop{12}{10}{3}t)\exp(\alpha_6\Bop{02}{01}{3}t)
   \nonumber \\
  \fl \cdot \exp[(\alpha_3^{21}\Lop{S}_3^{21}
  + \alpha_3^{20}\Lop{S}_3^{20} + \alpha_{-}^{21}\Lop{S}_{-}^{21} +
  \alpha_{-}^{20}\Lop{S}_{-}^{20} + \alpha_{+}^{21}\Lop{S}_{+}^{21} + \alpha_{+}^{20}\Lop{S}_{+}^{20})t]\Lket{\StateOp(0)},
\end{eqnarray}
where $\alpha_1,\ldots,\alpha_6$ are the coefficients of the
corresponding superoperators in
(\ref{eq:master_equation_superoperators}). Since the superoperators in
(\ref{eq:superoperators_sl3}) --excluding $\Lop{S}^{10}_{3}$-- are a
realization of the generators of $\sln{3}$, we may disentangle the
last exponential in (\ref{eq:master_equation_solution}) as
\begin{eqnarray}
  \label{eq:sl3_disentangle}
  \fl \exp\Bigl[\Bigl(\sum_{\ell_i}\alpha^{\ell_i}_{+}\Lop{S}^{\ell_i}_{+} +
  \sum_{\ell_j}\alpha^{\ell_j}_3\Lop{S}_3^{\ell_j} + \sum_{\ell_k} \alpha^{\ell_k}_{-}\Lop{S}^{\ell_k}_{-}
  \Bigr) t \Bigr] \nonumber \\
  = \prod_{\ell_i}\exp(\beta^{\ell_i}_{+}(t)\Lop{S}^{\ell_i}_{+})\prod_{\ell_j}\exp(\beta^{\ell_j}_{3}(t)\Lop{S}^{\ell_j}_3)\prod_{\ell_k}\exp(\beta^{\ell_k}_{-}(t)\Lop{S}^{\ell_k}_{-}),
\end{eqnarray}
where $\ell_i \in (20,10,21)$, $\ell_j \in (21,20)$ and
$\ell_k \in (21,10,20)$. The corresponding BCH formulas are obtained
calculating the exponentials in (\ref{eq:sl3_disentangle}) (using the
$3 \times 3$ matrix representation of $\sln{3}$) and solving a system of
nonlinear equations:
\begin{eqnarray}
  \label{eq:BCH_sl3}
  \fl e^{- \beta^{21}_{3}(t)} =
  -e^{-\beta^{20}_{3}(t)}\beta^{10}_{+}(t) \beta^{10}_{-}(t) +
  \frac{1}{D}\left[ - \alpha^{21}_{3} f_{1}(t) + f_{0}(t) + f_{2}(t)
  \left(\alpha^{21}_{+} \alpha^{21}_{-} + (\alpha^{21}_{3})^{2}\right)\right],\nonumber\\
  \fl e^{- \beta^{20}_{3}(t)} = \frac{1}{D} \left[- \alpha^{20}_{3} f_{1}(t) +
  f_{0}(t) + f_{2}(t) \left(\alpha^{20}_{+} \alpha^{20}_{-} +
  (\alpha^{20}_{3})^{2}\right)\right],\nonumber\\
 \fl \beta^{21}_{+}(t) = e^{\beta^{21}_{3}(t)}\left\{- \beta^{10}_{-}(t) \beta^{20}_{+}(t) e^{- \beta^{20}_{3}(t)} + \frac{\alpha^{21}_{+}}{D} \left[\alpha^{20}_{3}
  f_{2}(t) + f_{1}(t)\right]\right\},\nonumber\\
  \fl \beta^{21}_{-}(t) = e^{\beta^{21}_{3}(t)} \left\{- \beta^{10}_{+}(t) \beta^{20}_{-}(t) e^{- \beta^{20}_{3}(t)} + \frac{\alpha^{21}_{-}}{D} \left[\alpha^{20}_{3}
  f_{2}(t) + f_{1}(t)\right]\right\},\\
  \fl \beta^{10}_{+}(t) = \frac{\alpha^{20}_{+} f_{2}(t)}{D} \alpha^{21}_{-}
  e^{\beta^{20}_{3}(t)}, \qquad\qquad\qquad \beta^{20}_{+}(t) = \frac{\alpha^{20}_{+} e^{\beta^{20}_{3}(t)}}{D}
  \left[\alpha^{21}_{3} f_{2}(t) + f_{1}(t)\right],\nonumber\\
   \fl \beta^{10}_{-}(t) = \frac{\alpha^{20}_{-} f_{2}(t)}{D} \alpha^{21}_{+}
  e^{\beta^{20}_{3}(t)},\qquad\qquad\qquad
 \beta^{20}_{-}(t) = \frac{\alpha^{20}_{-} e^{\beta^{20}_{3}(t)}}{D}
  \left[\alpha^{21}_{3} f_{2}(t) + f_{1}(t)\right],\nonumber
\end{eqnarray}
with ($\lambda,\mu,\nu$ are the eigenvalues of the matrix in the
argument of the exponential in the l.h.s of
(\ref{eq:sl3_disentangle}))
\begin{eqnarray}
  \label{eq:D_f0_f1_f2_N0}
  D = (\mu - \lambda)(\nu - \lambda)(\mu - \nu),\nonumber\\
  f_0(t) = (\mu^2\nu - \mu\nu^2)e^{\lambda t} + (\nu^2\lambda - \nu\lambda^2)e^{\mu t} + (\lambda^2\mu - \lambda\mu^2)e^{\nu t},\\ 
  f_1(t) = (\nu^2 - \mu^2)e^{\lambda t} + (\lambda^2 - \nu^2)e^{\mu t} + (\mu^2 - \lambda^2)e^{\nu t},\nonumber\\
  f_2(t) = (\mu - \nu)e^{\lambda t} + (\nu - \lambda)e^{\mu t} + (\lambda - \mu)e^{\nu t}.\nonumber
\end{eqnarray}
If the radiation bath is in the ground state ($N_0 = 0$, $\mu = \nu$),
\begin{equation*}
  \label{eq:D_f0_f1_f2_ground_state}
  D = \lambda - \mu,\quad f_0(t) = \lambda e^{\mu t} - \mu e^{\lambda t},\quad 
  f_1(t) = e^{\lambda t} - e^{\mu t},\quad f_2(t) = 0,
\end{equation*}
and the BCH relations simplify considerably:
\begin{eqnarray}
  \label{eq:BCH_ground_state}
  \beta_3^{20}(t) = \beta_3^{21}(t) = -2\gamma t/3, \qquad\qquad
  \beta_{+}^{21} = \beta_{+}^{20} = \beta_{\pm}^{10} = 0, \nonumber\\
  \beta_{-}^{21}(t) = \frac{\gamma_{21}}{\gamma}(1-\e^{-\gamma t}),\qquad\qquad
  \beta_{-}^{20}(t) = \frac{\gamma_{20}}{\gamma}(1-\e^{-\gamma t}).
\end{eqnarray}

Since an arbitrary, symmetric initial state operator is a linear
combination of symmetric vectors (\ref{eq:symmetric_state_operator}),
in order to obtain an expression for $\Lket{\StateOp(t)}$ we need to
calculate:
\begin{eqnarray}
  \label{eq:evolution_symmetric_vector}
  \fl \e^{\Lop{L}t}|S_{n_k}) =
  \sum_{i_{+} = 0}^{n_{00}}{ n_{00} + i_{-} + j_{-} - j_{+} \choose i_{+}}
  (\beta_{+}^{20})^{i_{+}}\sum_{j_{+} = 0}^{n_{00}}{ n_{00} + i_{-} +
    j_{-} \choose
    j_{+}} (\beta_{+}^{10})^{j_{+}}\nonumber\\
  \fl \times \sum_{k_{+} = 0}^{n_{11}}{ n_{11} -
    j_{-} + k_{-}
    \choose k_{+}}
  (\beta_{+}^{21})^{k_{+}}\exp(\beta_3^{21}\tilde{a}^{21}_3) \exp(\beta_3^{20}\tilde{a}^{20}_3)\sum_{k_{-} = 0}^{n_{22}}{
    n_{22} - i_{-} \choose k_{-}} (\beta_{-}^{21})^{k_{-}}\nonumber\\
  \fl \times \sum_{j_{-} =
    0}^{n_{11}}{ n_{11} \choose j_{-}}
  (\beta_{-}^{10})^{j_{-}}\sum_{i_{-} = 0}^{n_{22}}{ n_{22} \choose
    i_{-}} (\beta_{-}^{20})^{i_{-}} C(t)
\raisebox{-3pt}{\Huge{Q}}\arraycolsep=1.5pt%
  \scriptsize{\begin{array}{lll}
     \nu_{00}  & n_{01} & \\
     n_{10}  & \nu_{11} & n_{12}\\
     n_{20}  & n_{21} & \nu_{22}
  \end{array}}\normalsize,
\end{eqnarray}
where
\begin{eqnarray}
  \label{eq:abbreviations}
  \fl C(t) = e^{\alpha_1t}\exp[\tfrac12\alpha_6(N-\nu_{00}
  - \nu_{11} - \nu_{22} - n_{10} - n_{20} - n_{21} -
  n_{12} - 2n_{01})t] \times\nonumber\\ \fl\exp[\tfrac12\alpha_2(n_{21}-n_{12})t]\exp[\tfrac12\alpha_3(n_{21}-n_{01})t]\exp(\tfrac12\alpha_4(n_{20}-n_{10})t]\exp[\tfrac12\alpha_5(n_{12}-n_{10})t],\nonumber\\
  \fl \tilde{a}^{21}_3 = [n_{22} - (n_{11} - j_{-} + k_{-})]/2,\nonumber\\
  \fl \tilde{a}^{20}_3 = [n_{22} - (n_{00} + i_{-} + j_{-})]/2,\\
  \fl \nu_{00} = n_{00} + i_{-} + j_{-} - j_{+} - i_{+} \nonumber\\
  \fl \nu_{11} = n_{11} - j_{-} + k_{-} - k_{+} + j_{+} \nonumber\\
  \fl \nu_{22} = n_{22} - i_{-} - k_{-} + k_{+} + i_{+} \nonumber.
\end{eqnarray}

In summary, the analytic solution to the Lindblad master equation
(\ref{eq:master_eq_lambda}) is given by the expression:
\begin{eqnarray}
  \label{eq:solution_master_equation}
  \fl  \Lket{\StateOp(t)} &= \e^{\alpha_1t}\exp(\alpha_2\Bop{21}{12}{3}t)\exp(\alpha_3\Bop{21}{01}{3}t)\exp(\alpha_4\Bop{20}{10}{3}t)\exp(\alpha_5\Bop{12}{10}{3}t)\exp(\alpha_6\Bop{02}{01}{3}t)
   \nonumber \\
  \fl & \times \exp(\beta_3^{21}(t)\Lop{S}_3^{21})
  \exp(\beta_3^{20}(t)\Lop{S}_3^{20}) \exp(\beta_{-}^{21}(t)\Lop{S}_{-}^{21})
  \exp(\beta_{-}^{20}(t)\Lop{S}_{-}^{20})
        \exp(\beta(t)_{+}^{21}\Lop{S}_{+}^{21}) \nonumber\\  
\fl & \times \exp(\beta_{+}^{20}(t)\Lop{S}_{+}^{20})\Lket{\StateOp(0)},
\end{eqnarray}
where the betas were calculated in (\ref{eq:BCH_sl3}) and the explicit
expression for $|\StateOp(t))$ can be obtained evaluating
(\ref{eq:evolution_symmetric_vector}) for each of the component vectors
of the initial state operator. To the best of our knowledge, this
solution has not been obtained before.

It is worth pointing out that even though the superoperators
describing the dynamics of a three-level system are generators of
$\sln{9}$, in many situations it is only necessary to disentangle an
exponential of elements of a lower-dimensional algebra in order to
obtain an analytic expression for $\StateOp(t)$. In the example above
we disentangled an exponential of elements of sl(3).

In addition to the results presented above, disentangling
(\ref{eq:master_equation_solution}) allows studying the dynamics of
the open system in a basis of coherent states in Liouville space. This
has been done recently for one two-level system interacting with a
bosonic bath \cite{ringel_liouville_2012}. Working in the Heisenberg
picture, the same procedure allows finding conserved quantities of the
open system \cite{albert_symmetries_2014}.

As a final remark, we would like to point out that in the case that
the system parameters in the Lindblad equation are time-dependent, the
tools developed here, namely the bosonized superoperators and the
symmetric basis vectors, together with the Lie-algebraic method
discussed in \cite{Jie1997} enable finding a semi-analytic solution,
that would involve numerically integrating a system of differential
equations. Additionally, it would be feasible to study how the system
approaches its equilibrium state.



\section{Conclusions}
\label{sec:conclusions}

In this work we considered the permutation-symmetric Lindblad equation
describing the open-system dynamics of a collection of $N$ $M$-level
systems subject to independent dissipation processes. We constructed a
basis of the symmetric subspace of Liouville space and showed that its
dimension grows polynomially with the number of systems. Therefore,
given a symmetric initial state operator, this result can be used to
efficiently simulate any symmetric Lindblad equation, since the
computational resources required are substantially reduced compared to
a simulation in the whole Liouville space. We also built a set of
superoperators, whose action on this basis is easily specified, that
are generators of the Lie algebra $\sln{M^2}$ and thus enable writing
any such Lindblad equation. We showed that these results can be used
to obtain an analytic solution of the Lindblad equation by means of
Lie-algebraic methods.

In order to show the usefulness of these results, we calculated the
evolution of the state operator of a collection of three-level atoms
interacting with independent radiation baths. Moreover, even though
the master equation is written in terms of elements of $\sln{9}$,
finding the solution required working only with an $\sln{3}$
subalgebra.



\ack

This work was supported by DGAPA-UNAM under grant PAPIIT IN103714.


\appendix

\section{Bosonized collective superoperators, generators of $\sln{9}$}
\label{sec:superops_generators_sl9}

The following table enables a researcher to quickly rewrite a master
equation in a form suitable for application of the solution method
outlined in section 5.

\arraycolsep=2pt
\[
\fl
\begin{array}{*{18}l}
\sum_i\hat{I}_{{3}}^{(i)} \hat{\rho} \hat{\sigma}^{{10(i)}}_{{\pm}} &=& (\Bop{11}{10}{\mp} + \Bop{01}{00}{\mp} + \Bop{21}{20}{\mp}) \hat{\rho} &&&&&&&&&&&&&\sum_i \Hop{\sigma}^{{21(i)}}_{{\pm}} \StateOp \Hop{\sigma}^{{21(i)}}_{{\mp}}&=&\Bop{22}{11}{\pm} \StateOp  \\[1.5ex] 
\sum_i\hat{\sigma}^{{10(i)}}_{{\pm}} \hat{\rho} \hat{I}_{{3}}^{(i)} &=& (\Bop{12}{02}{\pm} + \Bop{11}{01}{\pm} + \Bop{10}{00}{\pm}) \hat{\rho} &&&&&&&&&&&&&\sum_i \Hop{\sigma}^{{21(i)}}_{{\pm}} \StateOp \Hop{\sigma}^{{20(i)}}_{{\mp}}&=&\Bop{22}{10}{\pm} \StateOp  \\[1.5ex] 
\sum_i\hat{I}_{{3}}^{(i)} \hat{\rho} \hat{\sigma}^{{21(i)}}_{{\pm}} &=& (\Bop{22}{21}{\mp} + \Bop{12}{11}{\mp} + \Bop{02}{01}{\mp}) \hat{\rho} &&&&&&&&&&&&&\sum_i \Hop{\sigma}^{{20(i)}}_{{\pm}} \StateOp \Hop{\sigma}^{{21(i)}}_{{\mp}}&=&\Bop{22}{01}{\pm} \StateOp  \\[1.5ex]
\sum_i\hat{\sigma}^{{21(i)}}_{{\pm}} \hat{\rho} \hat{I}_{{3}}^{(i)} &=& (\Bop{21}{11}{\pm} + \Bop{20}{10}{\pm} + \Bop{22}{12}{\pm}) \hat{\rho} &&&&&&&&&&&&&\sum_i \Hop{\sigma}^{{20(i)}}_{{\pm}} \StateOp \Hop{\sigma}^{{20(i)}}_{{\mp}}&=& \Bop{22}{00}{\pm} \StateOp \\[1.5ex] 
\sum_i\hat{\sigma}^{{20(i)}}_{{\pm}} \hat{\rho} \hat{I}_{{3}}^{(i)} &=& (\Bop{21}{01}{\pm} + \Bop{20}{00}{\pm} + \Bop{22}{02}{\pm}) \hat{\rho} &&&&&&&&&&&&&\sum_i \Hop{\sigma}^{{21(i)}}_{{\pm}} \StateOp \Hop{\sigma}^{{21(i)}}_{{\pm}}&=& \Bop{21}{12}{\pm} \StateOp \\[1.5ex] 
\sum_i\hat{I}_{{3}}^{(i)} \hat{\rho} \hat{\sigma}^{{20(i)}}_{{\pm}} &=& (\Bop{22}{20}{\mp} + \Bop{12}{10}{\mp} + \Bop{02}{00}{\mp}) \hat{\rho} &&&&&&&&&&&&&\sum_i \Hop{\sigma}^{{21(i)}}_{{\pm}} \StateOp \Hop{\sigma}^{{10(i)}}_{{\mp}}&=& \Bop{21}{10}{\pm} \StateOp \\[1.5ex]   
  \sum_i\hat{\sigma}^{{20(i)}}_{{\pm}} \hat{\rho}
  \hat{\sigma}^{{02(i)}}_{{3}} &=&  \frac12(\Bop{22}{02}{\pm} - \Bop{20}{00}{\pm}) \hat{\rho} &&&&&&&&&&&&&\sum_i \Hop{\sigma}^{{20(i)}}_{{\pm}} \StateOp \Hop{\sigma}^{{21(i)}}_{{\pm}}&=& \Bop{21}{02}{\pm} \StateOp \\[1.5ex] 
\sum_i\hat{\sigma}^{{02(i)}}_{{3}} \hat{\rho} \hat{\sigma}^{{20(i)}}_{{\pm}} &=& \frac12(\Bop{22}{20}{\mp} - \Bop{02}{00}{\mp}) \hat{\rho} &&&&&&&&&&&&&\sum_i \Hop{\sigma}^{{20(i)}}_{{\pm}} \StateOp \Hop{\sigma}^{{10(i)}}_{{\mp}}&=& \Bop{21}{00}{\pm} \StateOp \\[1.5ex]
  \sum_i\hat{\sigma}^{{21(i)}}_{{\pm}} \hat{\rho}
  \hat{\sigma}^{{02(i)}}_{{3}} &=&  \frac12(\Bop{22}{12}{\pm} - \Bop{20}{10}{\pm}) \hat{\rho} &&&&&&&&&&&&&\sum_i \Hop{\sigma}^{{21(i)}}_{{\pm}} \StateOp \Hop{\sigma}^{{20(i)}}_{{\pm}}&=& \Bop{20}{12}{\pm} \StateOp \\[1.5ex] 
\sum_i\hat{\sigma}^{{02(i)}}_{{3}} \hat{\rho} \hat{\sigma}^{{21(i)}}_{{\pm}} &=& \frac12(\Bop{22}{21}{\mp} - \Bop{02}{01}{\mp}) \hat{\rho} &&&&&&&&&&&&&\sum_i \Hop{\sigma}^{{21(i)}}_{{\pm}} \StateOp \Hop{\sigma}^{{10(i)}}_{{\pm}}&=&\Bop{20}{11}{\pm} \StateOp  \\[1.5ex] 
\end{array}
\]

\[
\fl
\begin{array}{*{24}l}
\sum_i\hat{\sigma}^{{10(i)}}_{{\pm}} \hat{\rho} \hat{\sigma}^{{02(i)}}_{{3}} &=& \frac12(\Bop{12}{02}{\pm} - \Bop{10}{00}{\pm}) \hat{\rho} &&&&&&&&&&&&&&&&&&&\sum_i \Hop{\sigma}^{{20(i)}}_{{\pm}} \StateOp \Hop{\sigma}^{{20(i)}}_{{\pm}}&=& \Bop{20}{02}{\pm} \StateOp \\[1.5ex] 
\sum_i\hat{\sigma}^{{02(i)}}_{{3}} \hat{\rho}
  \hat{\sigma}^{{10(i)}}_{{\pm}} &=&  \frac12(\Bop{21}{20}{\mp} - \Bop{01}{00}{\mp}) \hat{\rho} &&&&&&&&&&&&&&&&&&&\sum_i \Hop{\sigma}^{{20(i)}}_{{\pm}} \StateOp \Hop{\sigma}^{{10(i)}}_{{\pm}}&=& \Bop{20}{01}{\pm} \StateOp \\[1.5ex]
\sum_i\hat{\sigma}^{{10(i)}}_{{\pm}} \hat{\rho} \hat{\sigma}^{{12(i)}}_{{3}} &=& \frac12(\Bop{12}{02}{\pm} - \Bop{11}{01}{\pm}) \hat{\rho} &&&&&&&&&&&&&&&&&&&\sum_i \Hop{\sigma}^{{10(i)}}_{{\pm}} \StateOp \Hop{\sigma}^{{21(i)}}_{{\mp}}&=&\Bop{12}{01}{\pm} \StateOp  \\[1.5ex] 
\sum_i\hat{\sigma}^{{12(i)}}_{{3}} \hat{\rho}
  \hat{\sigma}^{{10(i)}}_{{\pm}} &=&  \frac12(\Bop{21}{20}{\mp} - \Bop{11}{10}{\mp}) \hat{\rho} &&&&&&&&&&&&&&&&&&&\sum_i \Hop{\sigma}^{{10(i)}}_{{\pm}} \StateOp \Hop{\sigma}^{{20(i)}}_{{\mp}}&=&  \Bop{12}{00}{\pm} \StateOp\\[1.5ex]
\sum_i\hat{\sigma}^{{12(i)}}_{{3}} \hat{\rho} \hat{\sigma}^{{20(i)}}_{{\pm}} &=& \frac12(\Bop{22}{20}{\mp} - \Bop{12}{10}{\mp}) \hat{\rho} &&&&&&&&&&&&&&&&&&&\sum_i \Hop{\sigma}^{{10(i)}}_{{\pm}} \StateOp \Hop{\sigma}^{{21(i)}}_{{\pm}}&=&  \Bop{11}{02}{\pm} \StateOp\\[1.5ex] 
\sum_i\hat{\sigma}^{{20(i)}}_{{\pm}} \hat{\rho}
  \hat{\sigma}^{{12(i)}}_{{3}} &=&  \frac12(\Bop{22}{02}{\pm} - \Bop{21}{01}{\pm})\hat{\rho} &&&&&&&&&&&&&&&&&&&\sum_i \Hop{\sigma}^{{10(i)}}_{{\pm}} \StateOp \Hop{\sigma}^{{10(i)}}_{{\mp}}&=&  \Bop{11}{00}{\pm} \StateOp\\[1.5ex]
\sum_i\hat{\sigma}^{{12(i)}}_{{3}} \hat{\rho} \hat{\sigma}^{{21(i)}}_{{\pm}} &=& \frac12(\Bop{22}{21}{\mp} - \Bop{12}{11}{\mp}) \hat{\rho} &&&&&&&&&&&&&&&&&&&\,\sum_i \Hop{\sigma}^{{10(i)}}_{{\pm}} \StateOp \Hop{\sigma}^{{20(i)}}_{{\pm}}&=&  \Bop{10}{02}{\pm} \StateOp\\[1.5ex] 
 \sum_i\hat{\sigma}^{{21(i)}}_{{\pm}} \hat{\rho}
  \hat{\sigma}^{{12(i)}}_{{3}} &=& \frac12(\Bop{22}{12}{\pm} - \Bop{21}{11}{\pm}) \hat{\rho} &&&&&&&&&&&&&&&&&&&\,\sum_i \Hop{\sigma}^{{10(i)}}_{{\pm}} \StateOp \Hop{\sigma}^{{10(i)}}_{{\pm}}&=&\Bop{10}{01}{\pm} \StateOp 
\end{array}
\]

\[
\fl
\begin{array}{ll}
\sum_i\hat{\sigma}^{{12(i)}}_{{3}} \hat{\rho} \hat{\sigma}^{{02(i)}}_{{3}} &= \frac12(-\Bop{21}{01}{3} - \Bop{20}{10}{3} + \Bop{11}{02}{3} + \Bop{02}{01}{3} + \Bop{22}{11}{3} + \Bop{21}{12}{3}) \hat{\rho}\\[2ex]

\sum_i\hat{I}^{(i)}_{{3}} \hat{\rho} \hat{\sigma}^{{02(i)}}_{{3}} &= (- \Bop{21}{01}{3} - \Bop{20}{10}{3} + 2 \Bop{12}{10}{3} + \Bop{02}{01}{3} + \Bop{22}{00}{3} + \Bop{21}{12}{3}) \hat{\rho}\\[2ex]

\sum_i\hat{\sigma}^{{02(i)}}_{{3}} \hat{\rho} \hat{\sigma}^{{12(i)}}_{{3}} &= \frac12(-\Bop{21}{01}{3} + \Bop{11}{02}{3} + \Bop{22}{11}{3}) \hat{\rho}\\[2ex]

\sum_i\hat{\sigma}^{{12(i)}}_{{3}} \hat{\rho} \hat{I}^{(i)}_{{3}} &= (\Bop{20}{10}{3} + \Bop{22}{11}{3} + \Bop{21}{12}{3}) \hat{\rho}\\[2ex]

\sum_i\hat{\sigma}^{{02(i)}}_{{3}} \hat{\rho} \hat{I}^{(i)}_{{3}} &= (2 \Bop{21}{01}{3} + \Bop{20}{10}{3} - \Bop{12}{10}{3} - \Bop{02}{01}{3} + \Bop{22}{00}{3} - \Bop{21}{12}{3}) \hat{\rho}\\[2ex]

\sum_i\hat{I}^{(i)}_{{3}} \hat{\rho} \hat{\sigma}^{{12(i)}}_{{3}} &= (\Bop{02}{01}{3} + \Bop{22}{11}{3} - \Bop{21}{12}{3}) \hat{\rho}\\[2ex]

\sum_i\hat{\sigma}^{{12(i)}}_{{3}} \hat{\rho} \hat{\sigma}^{{12(i)}}_{{3}} &= (- \Bop{21}{01}{3} + \Bop{11}{02}{3} + \Bop{02}{01}{3} + \frac12\Bop{22}{11}{3} + \frac12\Bop{21}{12}{3}) \hat{\rho}\\[2ex]

  \sum_i\hat{\sigma}^{{02(i)}}_{{3}} \hat{\rho}
  \hat{\sigma}^{{02(i)}}_{{3}} &= \frac12(\Bop{12}{10}{3} +
                                 \Bop{02}{01}{3} 
                                 +
                                 \Bop{21}{12}{3} - \Bop{21}{01}{3} -
                                 \Bop{20}{10}{3} - \Bop{22}{00}{3} 
                                 ) \hat{\rho} +  (\Bop{11}{02}{3} + \Bop{22}{11}{3}) \hat{\rho}
\end{array}
\]



\section*{References}

\bibliographystyle{iopart-num}
\bibliography{refs}


\end{document}